# Dynamics of platicons due to third-order dispersion

VALERY E. LOBANOV[1], ARTEM V. CHERENKOV[1,2], ARTEM E. SHITIKOV[1,2], IGOR A. BILENKO[1,2] AND MICHAEL L. GORODETSKY[1,2]

[1]*Russian Quantum Center, Skolkovo 143025, Russia*
[2]*Faculty of Physics, Lomonosov Moscow State University, Moscow 119991, Russia*

**Abstract:** Dynamics of platicons caused by the third-order dispersion is studied. It is shown that under the influence of the third-order dispersion platicons obtain angular velocity depending both on dispersion and on detuning value. A method of tuning of platicon associated optical frequency comb repetition rate is proposed.

PACS 42.65.Tg – Optical solitons; nonlinear guided waves.
PACS 42.60.Da – Resonators, cavities, amplifiers, arrays, and rings.
PACS 42.65.Ky – Frequency conversion; harmonic generation, including higher-order harmonic generation.
PACS 42.65.Hw – Phase conjugation; photorefractive and Kerr effects.

Near-IR and visible spectral ranges are of particular interest for many practical applications of optical frequency combs [1-3]. For example, the spectral region of 650-1000 nm is considered as an "optical tissue window", since light with a wavelength in this window has minimal optical absorption and relatively small scattering in biological tissues. In the last years, spectroscopic methods using frequency combs generated in optical resonators [4-7] have attracted constantly growing interest among the fundamental and applied research community [7-9]. Unfortunately, most optical materials used in microresonators have normal group velocity dispersion (GVD) in visible and near-IR ranges, preventing self-starting generation of mode-locked frequency combs, as well as attractive bright dissipative Kerr solitons based combs in optical microresonators [9-14]. In this way, the development of new methods of generation of mode-locked frequency combs for microresonators with normal dispersion and study of properties of such frequency combs is of significant practical importance [15].

To date frequency combs in normal GVD microresonators were studied theoretically [16–18] and also demonstrated experimentally in different materials [19-22]. It was shown also that microresonators with normal GVD may support mode-locked frequency combs in the form of the dark optical solitons [18, 23-25]. A novel type of solitonic pulses, self-starting "platicon", was predicted in microresonators with normal dispersion under condition of shifted pump mode frequency resonance [26]. In real microresonators, such frequency shift may occur either due to the normal mode coupling between different mode families [27-29] or as a result of self-injection locking [20]. It was demonstrated that it is possible to change the duration of generated platicons in a wide range varying the pump detuning. It was shown also that generation of platicons is significantly more efficient than the generation of bright soliton trains in microresonators for the same absolute value of anomalous GVD in terms of conversion of the c.w. power into the power of the comb [26, 30]. Platicon generation is possible as well in absence of the pump mode shift when bichromatic or amplitude-modulated pump is used [31]. This method is efficient if pump modulation frequency or frequency difference between two pump waves is close to free spectral range (FSR) of the microresonator. An important issue of investigations in the field of microresonator solitons is a study of the influence of the higher-order dispersion on the properties of solitons or interaction of solitons with dispersive waves. This problem was actively studied for bright solitons in microresonators with anomalous dispersion. It was shown, for example, that the process of the dispersive wave formation caused by the higher-order dispersion terms may expand the comb generation bandwidth into the normal dispersion regime [32-36]. It was also shown, that the formation of the dispersive wave leads to a shift of the soliton spectrum maximum from the pump frequency (spectral recoil), while the soliton displaces the dispersive wave spectral peak from the zero-dispersion point [36]. However, for normal dispersion regime, this issue is less studied. Dispersive wave emission from dark solitons was demonstrated in [37].

In our paper, we study the dynamics of self-starting platicons generated using pumped mode shift influenced when the third-order dispersion (TOD) is present. For analysis, we used both coupled mode approach [38,39] and the model based on Lugiato-Lefever equation (LLE) [40,41] modified to include mode interaction [23, 26]. Namely, additional phase shift defined by the frequency shift value was applied to the central mode in the frequency domain step of the split-step Fourier routine. Results obtained by two methods were found to be in good agreement.

Our numerical model is based on the system of dimensionless coupled nonlinear mode equations modified to take the shift of the pumped mode into account [26]:

$$\frac{\partial a_\mu}{\partial \tau} = -\left(1 + i\zeta_\mu\right) a_\mu + i \sum_{\mu' \leq \mu''} \left(2 - \delta_{\mu'\mu''}\right) a_{\mu'} a_{\mu''} a^*_{\mu'+\mu''-\mu} + \delta_{0\mu} f. \qquad (1)$$

This system in fact is just a discrete Fourier transform of the LLE. We consider Taylor expansion of the dispersion law modified to take the pumped mode frequency shift $\Delta$ into account: $\omega_\mu = \omega_0 - \delta_{0\mu}\Delta + D_1\mu - \frac{1}{2}D_2\mu^2 + \frac{1}{6}D_3\mu^3$, where $\omega_0$ is the unperturbed pumped mode frequency. Note, that there is a minus before the quadratic term since GVD is normal (we assume that $D_2 > 0$). Here $\delta_{\mu'\mu''}$ is the Kronecker delta, $a_\mu = A_\mu \sqrt{2g/\kappa} \exp\left[-i\left(\omega_\mu - \omega_p - \mu D_1\right)t\right]$ can be interpreted as the slowly varying amplitude of the comb modes for the mode frequency $\omega_\mu$, $\tau = \kappa t/2$ denotes the normalized time, $D_1 = 2\pi/T_R$ is FSR, $T_R \approx 2\pi Rc/n_0$ is the round-trip time, $R$ is the radius of the resonator, $n_0$ is the refractive index, $c$ is the speed of light in vacuum, $g = \dfrac{\hbar\omega_0^2 c n_2}{n_0^2 V_{\mathit{eff}}}$ is the nonlinear coupling coefficient, $V_{\mathit{eff}}$ is the effective mode volume, $n_2$ is the nonlinear refractive index, $\kappa = \dfrac{\omega_0}{Q} = \kappa_0 + \kappa_{\mathit{ext}}$ denotes the cavity decay rate as the sum of intrinsic decay rate $\kappa_0$ and coupling rate $\kappa_{\mathit{ext}}$, where $Q$ is the total quality factor, $\eta = \kappa_{\mathit{ext}}/\kappa$ is the coupling efficiency ($\eta = 1/2$ for the critical coupling), $f = \sqrt{\dfrac{8 g \eta P_0}{\kappa^2 \hbar \omega_0}}$ stands for the dimensionless pump amplitude. All mode numbers $\mu$ are defined relative to the pumped mode $\mu = m - m_0$ with the initial azimuthal number $m_0 \approx 2\pi R n_0/\lambda$, where $\lambda = 2\pi c/\omega_0$ is the wavelength. $\zeta_\mu = 2(\omega_\mu - \omega_p - \mu D_1)/\kappa$ stands for the normalized detuning.

The corresponding LLE equation is following:

$$\frac{\partial \psi}{\partial \tau} = -i\frac{1}{2}\beta_2 \frac{\partial^2 \psi}{\partial \varphi^2} + \frac{1}{6}\beta_3 \frac{\partial^3 \psi}{\partial \varphi^3} + i|\psi|^2 \psi - (1 + i\bar{\zeta}_0)\psi + f, \qquad (2)$$

where $\beta_2 = \dfrac{2D_2}{\kappa}$, $\beta_3 = \dfrac{2D_3}{\kappa}$, $\psi(\varphi) = \sum_\mu a_\mu \exp(i\mu\varphi)$, $\bar{\zeta}_0 = \zeta_0 + (2\Delta/\kappa)$. Additional phase shift defined by the frequency shift value $\Delta$ was applied to the central mode in the frequency domain step of the split-step Fourier routine.

At first, we studied the generation of platicons from the noise-like input by the frequency scan ($\zeta_0 = \zeta_0(0) + \alpha\tau$) for different values of the third-order dispersion. We set $\Delta = 2\kappa$, $f = 4$, $D_2/\kappa = 0.0025$. For analysis, we calculated field distribution evolution in temporal representation. As it is shown in Figure 1 platicons may be generated in the limited frequency range (limited range of normalized detuning $\zeta_0$) even if the TOD dispersion is present ($D_3 \neq 0$). Interestingly, minimal possible detuning value slightly depends on $D_3$. The presence of the TOD leads to the distortion of the dynamical profile. Some inclination of the dynamical profile may be observed and this inclination increases with the growth of $D_3$ (see Figs. 1a – 1c).

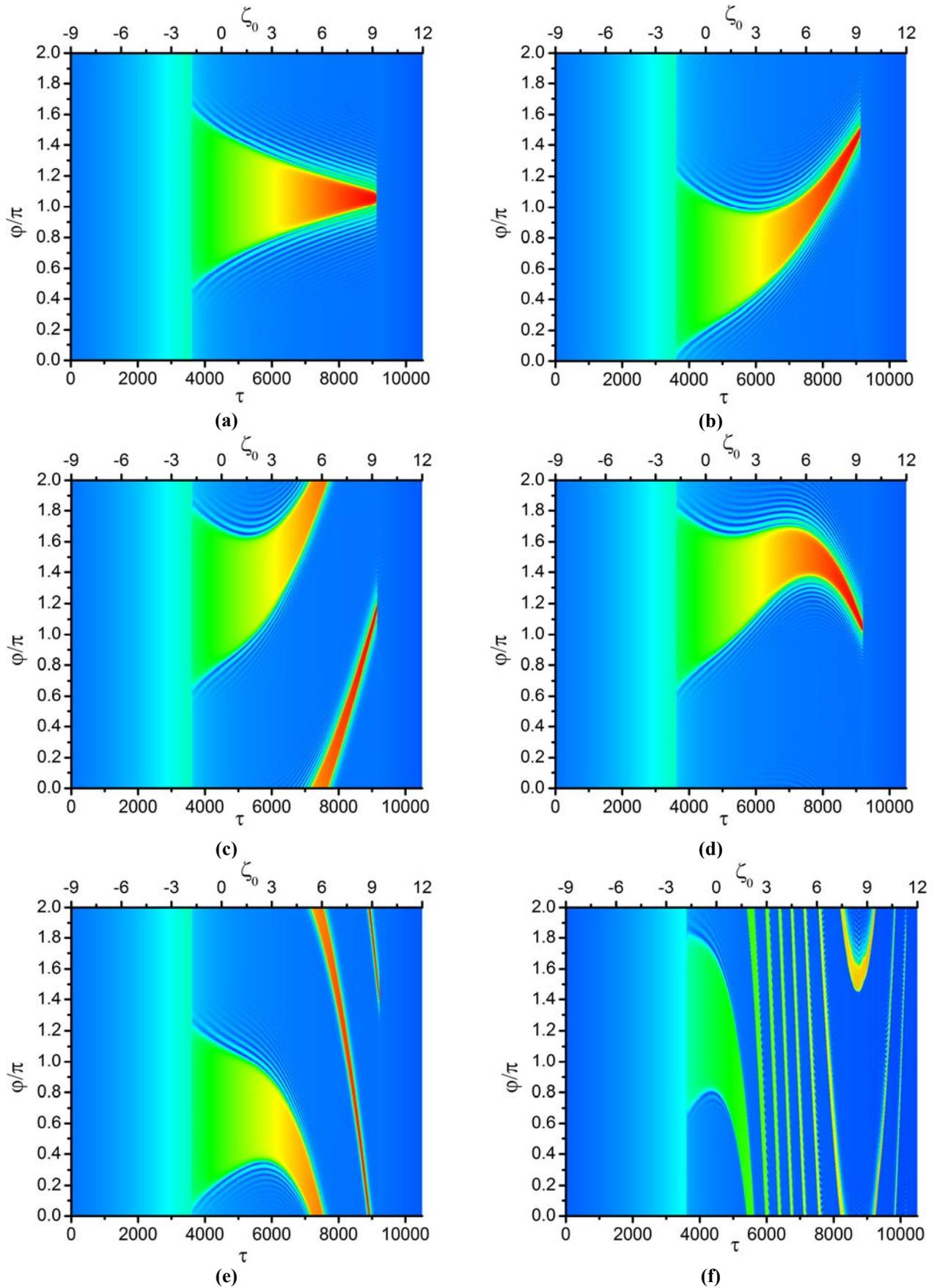

**Fig. 1.** Field distribution evolution upon frequency scan ($\zeta_0 = \zeta_0(0) + \alpha\tau$, $\alpha = 0.002$) in the temporal representation for different values of the third-order dispersion: **(a)** $D_3 = 0$; **(b)** $D_3/3D_2 = 0.0004$; **(c)** $D_3/3D_2 = 0.0012$; **(d)** $D_3/3D_2 = 0.003$; **(e)** $D_3/3D_2 = 0.004$; **(f)** $D_3/3D_2 = 0.008$.

Dynamical profile inclination is the result of the drift taking place due to the TOD [37]. However, at some $D_3$ value one may notice that the growth of inclination stops and then changes direction (Figs. 1d-1e). Further on, one more turning point may be found (see Fig. 1f).

One may conclude that the drift velocity depends both on mismatch value and the TOD value. Recently, it was found that for bright solitons $v_\varphi = \zeta_0 D_3 / 3D_2$ [36]. So, the drift direction for bright solitons is determined only by the sign of the TOD coefficient. Interestingly, in case of platicons with fixed sign of $D_3$ one may observe different drift directions depending on the absolute value of the TOD and detuning. To check this, we simulate platicon propagation at fixed mismatch value for different values of $D_3$ using platicon profiles obtained for $D_3 = 0$ as an input.

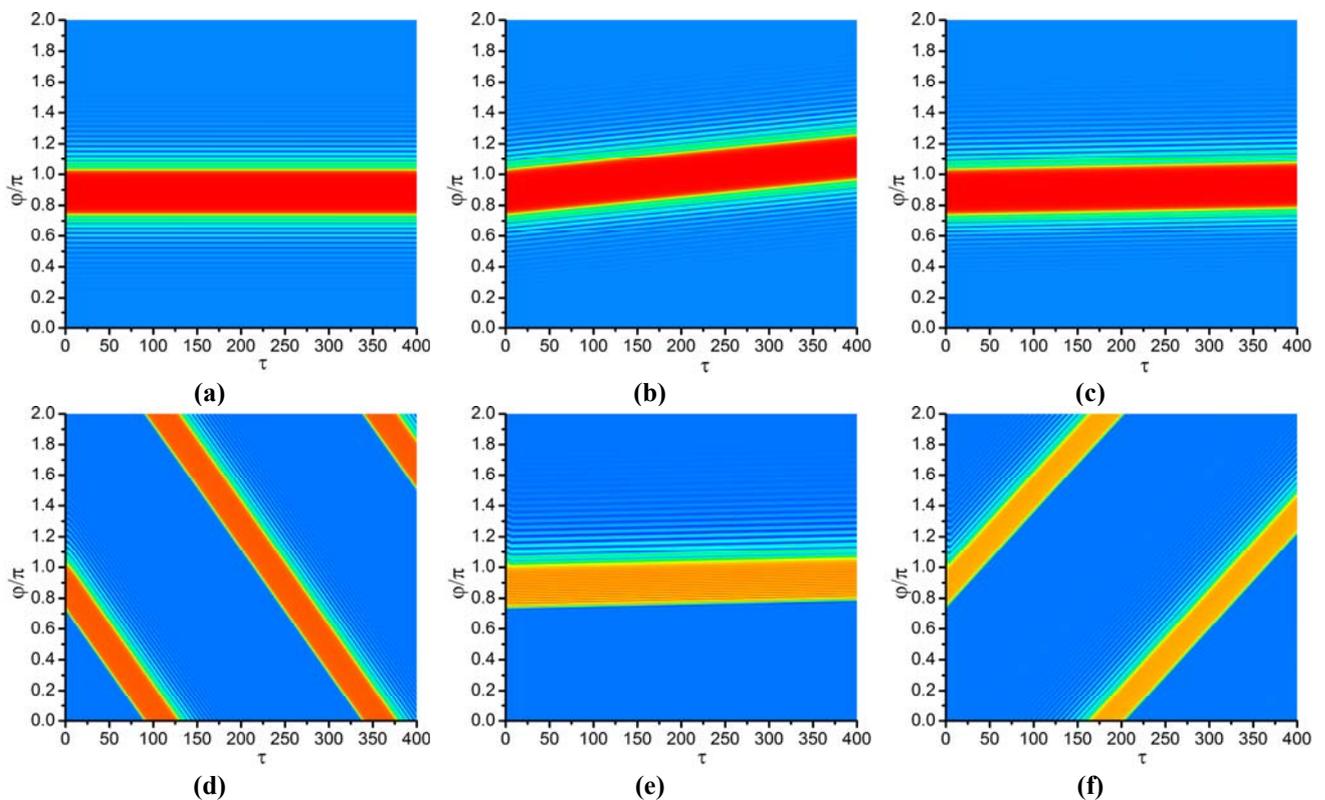

**Fig. 2. Platicon propagation at $\zeta_0 = 6$ for different values of the third-order dispersion: (a)** $D_3 = 0$; **(b)** $D_3 / 3D_2 = 0.0012$; **(c)** $D_3 / 3D_2 = 0.0028$; **(d)** $D_3 / 3D_2 = 0.0064$; **(e)** $D_3 / 3D_2 = 0.0096$; **(f)** $D_3 / 3D_2 = 0.012$.

As it is shown in Figure 2 platicons propagate in a stable manner having drift velocity depending on the TOD value. Also, with the growth of $D_3$, the platicon profile (flat-top with symmetric oscillating tails) becomes indented and asymmetric with a pronounced one-sided oscillating tail (see Figs. 2e-2f and Fig. 3).

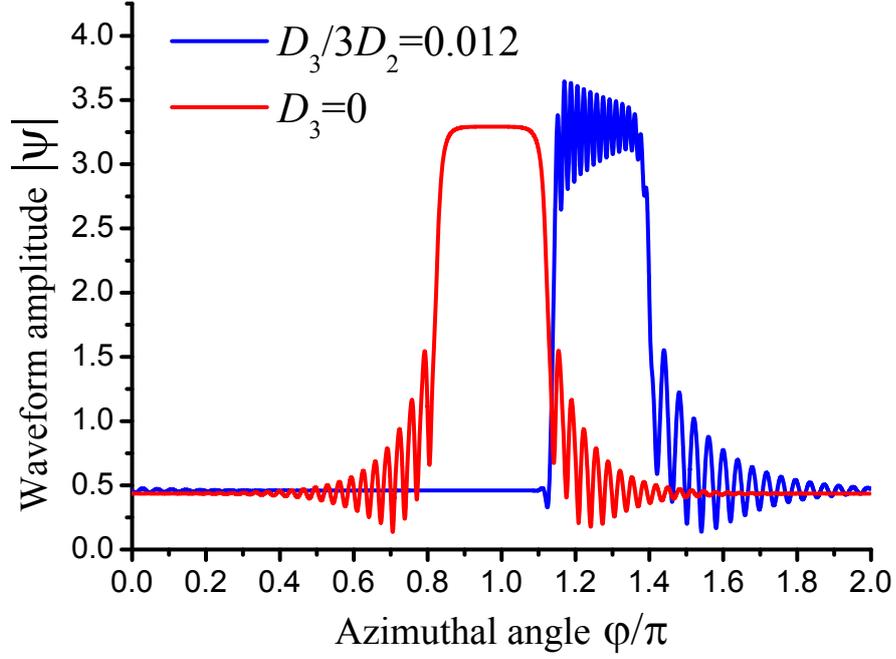

**Fig. 3.** Platicon profiles at $\zeta_0 = 6$ for different values of the third-order dispersion.

To define the velocity of the high-intensity part of the platicon we calculate the position of its center of mass upon propagation. However, since platicons have nonzero background and emit dispersive waves, it is inconvenient to use conventional definition $\varphi_c = \int_0^{2\pi} |\psi|^2 \varphi d\varphi / \int_0^{2\pi} |\psi|^2 d\varphi$. To minimize the influence of dispersive waves and background and to focus on the motion of the most intensive part of the platicon profile we define the position of a high-intensity part of platicon as following:

$$\overline{\varphi}_c = \int_0^{2\pi} |\psi|^6 \varphi d\varphi / \int_0^{2\pi} |\psi|^6 d\varphi. \qquad (3)$$

In Eq. 3 we used power value higher than 2 (namely 6) to select high-amplitude regions and to suppress low-amplitude background and found than power values higher than 6 provides practically the same result. Drift velocity was defined as

$$v_\varphi = \frac{d\overline{\varphi}_c}{d\tau}. \qquad (4)$$

As shown in Figure 4, the dependence of the drift velocity on the TOD value has a dip position and depth of which depend on the mismatch value. After this dip, the dependence is a monotonically growing function.

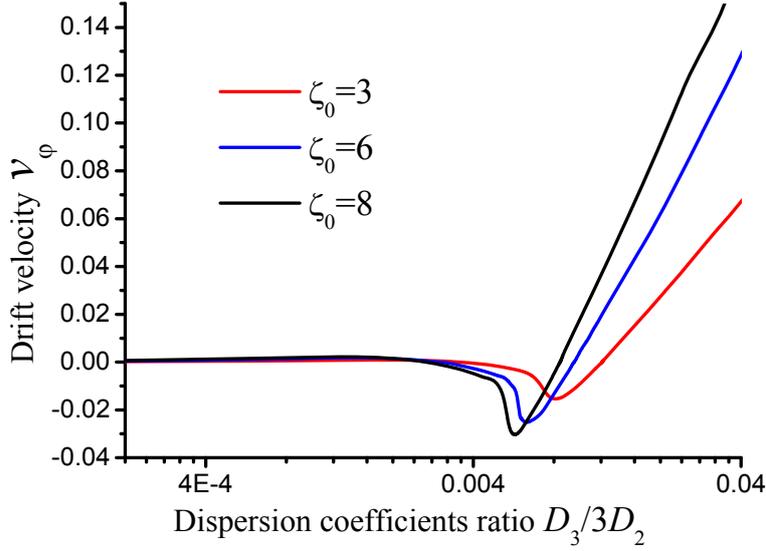

**Fig. 4.** Platicon drift velocity vs. $D_3$ for different values of $\zeta_0$.

Nonzero soliton velocity has an important consequence. Since the angular coordinate $\varphi$ is defined in a coordinate system rotating with angular velocity $D_1$ ($\varphi = \phi - D_1 t$, where $\phi$ is the regular polar angle inside the microresonator), the pulse velocity associated with the TOD may be interpreted as the shift of the pulse repetition rate (in absence of the third-order dispersion, the pulse repetition rate is equal to $D_1/2\pi$). In this way, the repetition rate $\nu_r$ may be tuned by varying the laser detuning $\zeta_0$ since $2\pi\Delta\nu_r = \frac{\kappa}{2}v_\varphi$ (see Fig. 5). This effect may be used in practical application of soliton comb based microwave oscillators.

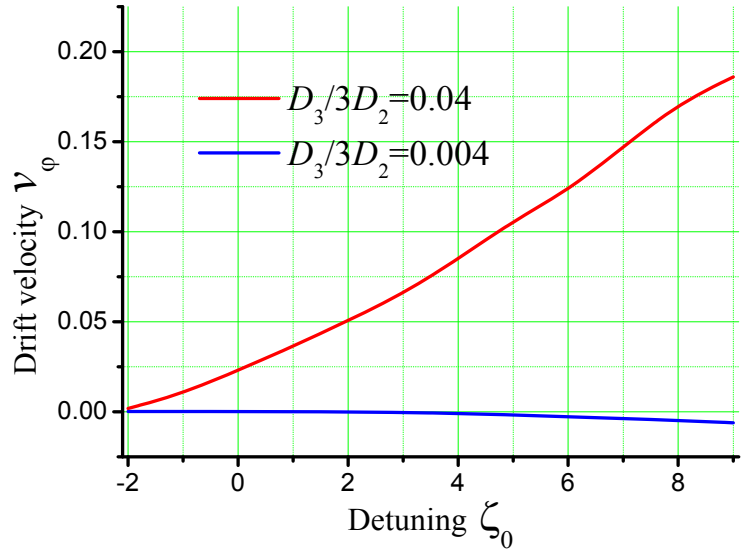

**Fig. 5.** Platicon drift velocity vs. $\zeta_0$ for different values of $D_3$.

Note, that dispersion parameters of crystalline microresonators can be specifically set, for example, by means of dispersion engineering [42-46]. In particular, it has been demonstrated that one may tune the effective dispersion of the microresonator varying its cross-sectional shape [45], e.g. by the formation of the precision micrometric protrusions [46].

To summarize, we have studied the dynamics of platicons caused by the third-order dispersion and have showed that under the influence of the third-order dispersion platicons obtain angular velocity depending both on dispersion and on detuning value. We have also proposed a method of tuning of platicon associated optical frequency comb repetition rate.

## ACKNOWLEDGEMENTS

This work was supported by the Russian Foundation for Basic Research (project 17-02-00522).

## AUTHOR CONTRIBUTION STATEMENT

V.E.L. developed the theory and performed the simulations. A.V.C. and A.E.S. assisted in simulations. V.E.L., I.A.B. and M.L.G. discussed all data in the manuscript. V.E.L. and M.LG. wrote the manuscript.